\pdfoutput=1          
\documentclass[11pt,a4paper]{article}

\usepackage[utf8]{inputenc}
\usepackage[T1]{fontenc}
\usepackage{lmodern}
\usepackage[left=2.5cm, right=2.5cm, top=2.5cm, bottom=2.5cm]{geometry}
\usepackage{amsmath,amssymb,amsfonts}
\usepackage{graphicx}
\usepackage{booktabs}
\usepackage{multirow}
\usepackage{xcolor}
\usepackage[numbers,sort&compress]{natbib}
\usepackage{hyperref}
\usepackage{cleveref}
\usepackage{caption}
\usepackage{subcaption}
\usepackage{float}
\usepackage{enumitem}
\usepackage{tikz}
\usetikzlibrary{shapes.geometric, arrows.meta, positioning, calc, fit}
\usepackage{abstract}

\usepackage{authblk}

\definecolor{qblue}{HTML}{2563EB}
\definecolor{qpurple}{HTML}{7C3AED}
\definecolor{accent}{HTML}{059669}

\hypersetup{
  colorlinks=true,
  linkcolor=qblue,
  citecolor=qpurple,
  urlcolor=accent
}

\title{\textbf{Hybrid Photonic Quantum Reservoir Computing\\for High-Dimensional Financial Surface Prediction}}

\author[1, 2, *]{Fyodor Amanov}
\author[1, 2, *]{Azamkhon Azamov}
\affil[1]{QuanTech}
\affil[2]{New Uzbekistan University}
\affil[*]{These authors contributed equally to this work.}
\date{March 2026}

\begin{document}
\maketitle

\begin{abstract}
We propose a hybrid photonic quantum reservoir computing (QRC) framework for swaption surface prediction. The pipeline compresses 224-dimensional surfaces to a 20-dimensional latent space via a sparse denoising autoencoder, extracts 1,215 Fock-basis features from an ensemble of three fixed photonic reservoirs, concatenates them with a 120-dimensional classical context, and maps the resulting 1,335-dimensional feature vector to predictions with Ridge regression.
We benchmark against 10 classical and quantum baselines on six held-out trading days. Our approach achieves the lowest surface RMSE of~$0.0425$ while maintaining sub-millisecond inference. The quantum layer has zero trainable parameters, sidestepping barren plateaus entirely. Variational quantum methods (VQC, Quantum LSTM) yield negative $R^{2}$ on test data, confirming that fixed quantum feature extractors paired with regularised readouts are more viable for low-data financial applications.

\end{abstract}

\section{Introduction}\label{sec:intro}

Financial derivatives pricing remains one of the most computationally demanding challenges in quantitative finance. Swaptions---options granting the right to enter interest rate swaps---are priced across two-dimensional grids of tenors and maturities, producing surfaces whose dynamics must be forecast day by day~\citep{Hull2018}. Each surface in this study contains 224 prices (a $14\!\times\!16$ grid), and only 494 historical trading days are available for training. This data regime sits in an uncomfortable gap: too high-dimensional for simple parametric models, yet too small for deep learning to generalise without aggressive regularisation.

Quantum computing has drawn attention as a potential tool for finance~\citep{Orus2019,Herman2023}, with applications from portfolio optimisation to Monte Carlo simulation~\citep{Stamatopoulos2020}. However, quantum machine learning (QML) for regression has seen mixed results~\citep{Schuld2019,Cerezo2021}. Variational quantum circuits (VQCs), while theoretically expressive~\citep{Abbas2021}, are plagued by barren plateaus---exponentially vanishing gradients that make training infeasible as circuit width grows~\citep{McClean2018}. On small datasets, parameter-heavy quantum models are particularly vulnerable to overfitting, a finding we confirm empirically.

Quantum reservoir computing (QRC) offers a fundamentally different paradigm. Instead of training quantum parameters via gradient descent, QRC uses fixed quantum systems as nonlinear feature extractors and trains only a classical readout~\citep{Fujii2017,Nakajima2019,Mujal2021}. This bypasses barren plateaus entirely, reduces training to a convex problem, and produces deterministic results. Photonic implementations are attractive because boson sampling---the physical process underlying Fock-state probability computation---is \#P-hard to simulate classically~\citep{Aaronson2011}, meaning photonic reservoirs produce features that classical computers cannot efficiently replicate.

Our contributions are:
\begin{enumerate}[nosep,leftmargin=*]
  \item A \textbf{three-stage robust preprocessing pipeline} tailored to fat-tailed financial data, fully invertible and free of temporal data leakage.
  \item A \textbf{sparse denoising autoencoder} with ELU bottleneck that compresses 224-dimensional surfaces to 20 latent dimensions.
  \item An \textbf{ensemble of three fixed photonic reservoirs} producing 1,215 Fock-basis features encoding pairwise, cubic, and quartic quantum correlations.
  \item A thorough \textbf{benchmark against 10 alternative models} spanning classical ML, deep learning, and variational quantum methods on a strictly held-out test set.
\end{enumerate}

\begin{figure}[t]
  \centering
  \resizebox{\linewidth}{!}{%
  \begin{tikzpicture}[
    node distance=0.7cm and 0.5cm,
    block/.style={rectangle, draw=qblue, fill=qblue!8, rounded corners=3pt,
                  minimum height=1.0cm, minimum width=2.0cm, align=center,
                  font=\small\sffamily},
    qblock/.style={rectangle, draw=qpurple, fill=qpurple!8, rounded corners=3pt,
                   minimum height=1.0cm, minimum width=2.0cm, align=center,
                   font=\small\sffamily},
    arr/.style={-{Stealth[length=5pt]}, thick, color=gray!70},
  ]
    \node[block] (raw) {Raw Swaption\\Surfaces\\$\mathbf{S}_t\!\in\!\mathbb{R}^{224}$};
    \node[block, right=of raw] (pre) {3-Stage\\Preprocessing\\Win$\to$Rob$\to$MM};
    \node[block, right=of pre] (ae)  {Sparse Denoising\\Autoencoder\\$224\!\to\!20$};
    \node[block, right=of ae]  (win) {Temporal\\Windowing\\$k\!=\!5$};
    \node[qblock, right=of win] (qrc) {Ensemble\\QORC\\$3\times$Reservoir};
    \node[block, right=of qrc] (cat) {Feature\\Concat\\$1335$-dim};
    \node[block, right=of cat] (rid) {Ridge\\Regression\\$\alpha\!=\!100$};
    \node[block, right=of rid] (dec) {AE Decoder\\$\hat{\mathbf{S}}_{t+1}\!\in\!\mathbb{R}^{224}$};

    \draw[arr] (raw) -- (pre);
    \draw[arr] (pre) -- (ae);
    \draw[arr] (ae)  -- (win);
    \draw[arr] (win) -- (qrc);
    \draw[arr] (qrc) -- (cat);
    \draw[arr] (win.south) -- ++(0,-0.4) -| (cat.south);
    \draw[arr] (cat) -- (rid);
    \draw[arr] (rid) -- (dec);
  \end{tikzpicture}%
  }
  \caption{End-to-end pipeline. Classical context (120-dim) from temporal windowing and quantum features (1,215-dim) from the QORC ensemble are concatenated before the Ridge readout. The autoencoder decoder reconstructs the full 224-dim surface. The quantum layer (purple) has zero trainable parameters.}
  \label{fig:pipeline}
\end{figure}
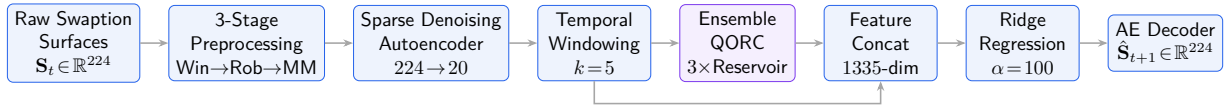

\section{Related Work}\label{sec:related}

\subsection{Quantum Reservoir Computing}
Quantum reservoir computing draws from classical echo state networks and liquid state machines~\citep{Jaeger2001,Maass2002}, extending them to quantum substrates. Fujii and Nakajima~\citep{Fujii2017} demonstrated that randomly coupled quantum spin systems can serve as universal computational reservoirs. Nakajima et~al.~\citep{Nakajima2019} provided early experimental evidence that fixed quantum dynamics generate computationally useful feature maps. Mart\'{i}nez-Pe\~{n}a et~al.~\citep{MartinezPena2021} studied memory and nonlinear processing capabilities, finding that quantum coherence contributes meaningfully to temporal processing. Mujal et~al.~\citep{Mujal2021} identified photonic platforms as promising candidates due to room-temperature operation and natural nonlinearity. More recently, Ghosh et~al.~\citep{Ghosh2021} formalised quantum reservoir processing theory, Suzuki et~al.~\citep{Suzuki2022} demonstrated QRC for temporal tasks, and Nokkala et~al.~\citep{Nokkala2024} proposed online learning extensions with lateral connections.

\subsection{Photonic Machine Learning}
Photonic systems offer inherent parallelism through superposition, natural nonlinearity through photon--photon correlations, and fast operation at optical clock speeds~\citep{Killoran2019}. Recent experimental progress in photonic quantum information processing~\citep{Spagnolo2022} further motivates photonic platforms. Output probabilities in the Fock basis depend on permanents of unitary submatrices---\#P-hard computations~\citep{Aaronson2011}. Perceval~\citep{Heurtel2023} and MerLin~\citep{Dudas2023} provide frameworks for constructing linear optical circuits. Our work uses the sandwich circuit architecture, which has been shown to produce rich mixing of input information.

\subsection{Quantum Methods in Finance}
Or\'{u}s et~al.~\citep{Orus2019} surveyed quantum computing for financial applications. Chakrabarti et~al.~\citep{Chakrabarti2021} applied variational methods to option pricing, and Brandhofer et~al.~\citep{Brandhofer2023} to portfolio optimisation, though practical advantages remain elusive. The key challenge we address is that financial datasets are small by ML standards---thousands of samples, not millions---making regularisation and model simplicity more important than expressiveness.

\section{Methodology}\label{sec:method}

\subsection{Problem Formulation}\label{sec:formulation}

Let $\mathbf{S}_t \in \mathbb{R}^{224}$ denote the vectorised swaption surface on trading day $t$, where $t = 1,\ldots,T$ and $T=494$. We compress each surface to a latent code $\mathbf{z}_t \in \mathbb{R}^{d}$ ($d\!=\!20$) via a learned encoder $f_\theta$, predict the next code $\hat{\mathbf{z}}_{t+1}$, and reconstruct through a decoder $g_\phi$:
\begin{equation}\label{eq:pipeline}
  \hat{\mathbf{S}}_{t+1} = g_\phi\!\left(\hat{\mathbf{z}}_{t+1}\right),
  \quad
  \hat{\mathbf{z}}_{t+1} = h\!\left(\mathbf{z}_{t-k+1},\ldots,\mathbf{z}_t\right).
\end{equation}

\subsection{Preprocessing}\label{sec:preprocessing}

Financial time series exhibit fat tails, heteroskedasticity, and extreme observations. We implement a three-stage pipeline that is robust, invertible, and free of look-ahead bias.

\paragraph{Stage 1: Winsorization.} For each of the 224 price dimensions, we clip values at the 1st and 99th percentiles computed on training data:
\begin{equation}\label{eq:winsorize}
  s_{t,j}^{(1)} = \text{clip}\!\left(s_{t,j},\; q_{0.01}^{(j)},\; q_{0.99}^{(j)}\right).
\end{equation}

\paragraph{Stage 2: Robust Scaling.} We centre by the median and scale by the interquartile range (IQR):
\begin{equation}\label{eq:robust}
  s_{t,j}^{(2)} = \frac{s_{t,j}^{(1)} - \mathrm{med}_j}{\mathrm{IQR}_j}\,,
\end{equation}
where $\mathrm{med}_j = Q_{0.5}^{(j)}$ and $\mathrm{IQR}_j = Q_{0.75}^{(j)} - Q_{0.25}^{(j)}$.

\paragraph{Stage 3: Min-Max Normalisation.} We rescale to $[0,1]$:
\begin{equation}\label{eq:minmax}
  s_{t,j}^{(3)} = \frac{s_{t,j}^{(2)} - \min_j}{\max_j - \min_j}\,.
\end{equation}
All statistics ($q_\alpha$, med, IQR, min, max) are fitted exclusively on training data and applied identically to validation and test data. The chain is fully invertible for back-transforming predictions to original prices.

\begin{figure}[t]
  \centering
  \includegraphics[width=\linewidth]{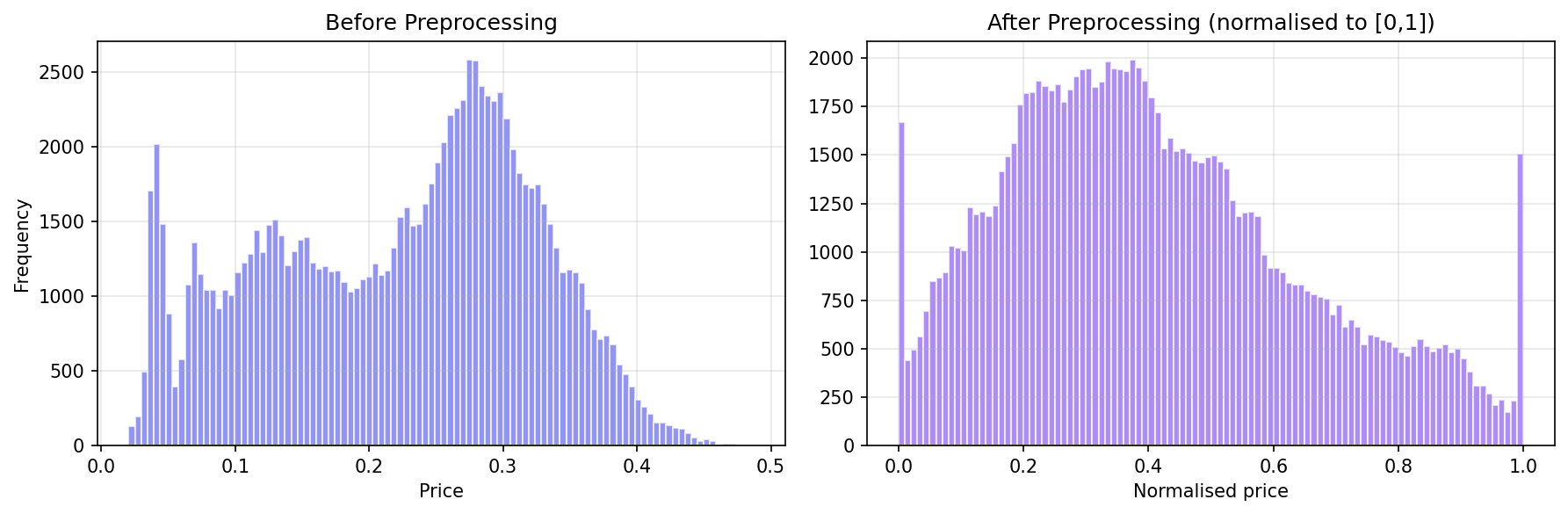}
  \caption{Effect of the three-stage preprocessing pipeline. Left: raw swaption distributions with extreme tails. Right: after Winsorize$\to$RobustScale$\to$MinMax, values lie cleanly in $[0,1]$.}
  \label{fig:preprocessing}
\end{figure}

\subsection{Sparse Denoising Autoencoder}\label{sec:autoencoder}

\subsubsection{Architecture}
The autoencoder uses a symmetric encoder--decoder design:
\begin{align}
  \text{Enc:}& \;\; \mathbb{R}^{224} \xrightarrow{\text{ReLU}} \mathbb{R}^{128} \xrightarrow{\text{ReLU}} \mathbb{R}^{64} \xrightarrow{\text{ELU}} \mathbb{R}^{20} \label{eq:encoder}\\
  \text{Dec:}& \;\; \mathbb{R}^{20} \xrightarrow{\text{ReLU}} \mathbb{R}^{64} \xrightarrow{\text{ReLU}} \mathbb{R}^{128} \xrightarrow{\sigma} \mathbb{R}^{224} \label{eq:decoder}
\end{align}
The bottleneck uses the Exponential Linear Unit:
\begin{equation}\label{eq:elu}
  \text{ELU}(x) = \begin{cases} x & x > 0 \\ e^x - 1 & x \leq 0 \end{cases}
\end{equation}
rather than ReLU, which can cause latent dimensions to ``die'' (become permanently zero) under L1 sparsity penalties.

\subsubsection{Training Objective}
The loss combines reconstruction fidelity with a sparsity penalty:
\begin{equation}\label{eq:ae_loss}
  \mathcal{L}_{\text{AE}} = \frac{1}{N}\sum_{t=1}^{N}\!\left\|\hat{\mathbf{S}}_t - \mathbf{S}_t\right\|_2^2 + \lambda \frac{1}{N}\sum_{t=1}^{N}\!\left\|\mathbf{z}_t\right\|_1
\end{equation}
with $\lambda = 10^{-4}$.

\subsubsection{Denoising}
During training, 15\% of input dimensions are randomly masked before encoding, while the reconstruction target remains the clean input:
\begin{equation}\label{eq:denoising}
  \tilde{\mathbf{S}}_t = \mathbf{S}_t \odot \mathbf{m}_t, \quad m_{t,j} \sim \text{Bernoulli}(0.85)
\end{equation}
This regularises the encoder and teaches robust reconstruction from partial observations.

\begin{figure}[t]
  \centering
  \includegraphics[width=\linewidth]{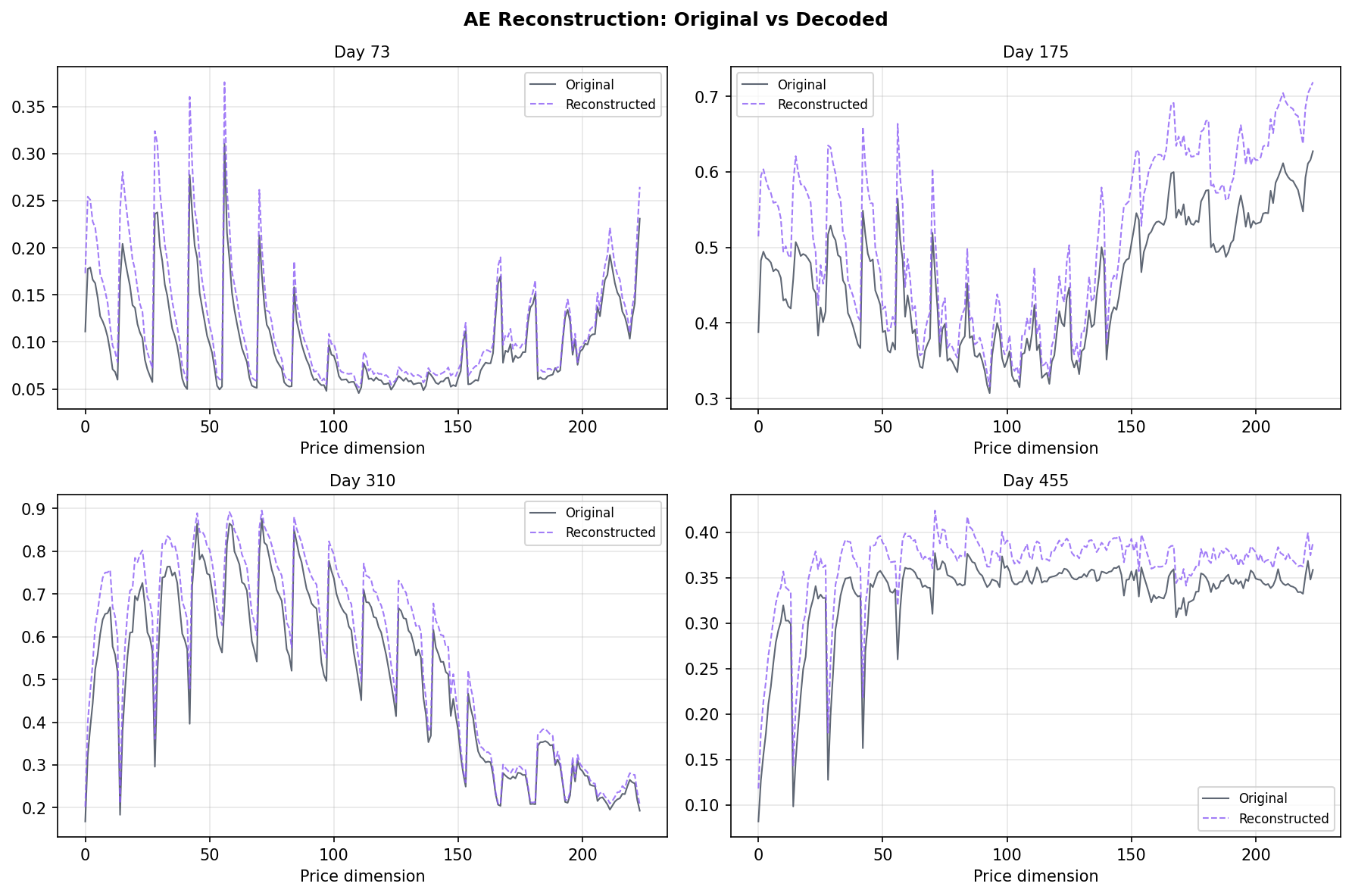}
  \caption{Autoencoder reconstruction quality. Original (top) vs.\ decoded (bottom) swaption surfaces for selected days, demonstrating high fidelity of the 224$\to$20$\to$224 compression.}
  \label{fig:ae_recon}
\end{figure}

\begin{figure}[t]
  \centering
  \includegraphics[width=\linewidth]{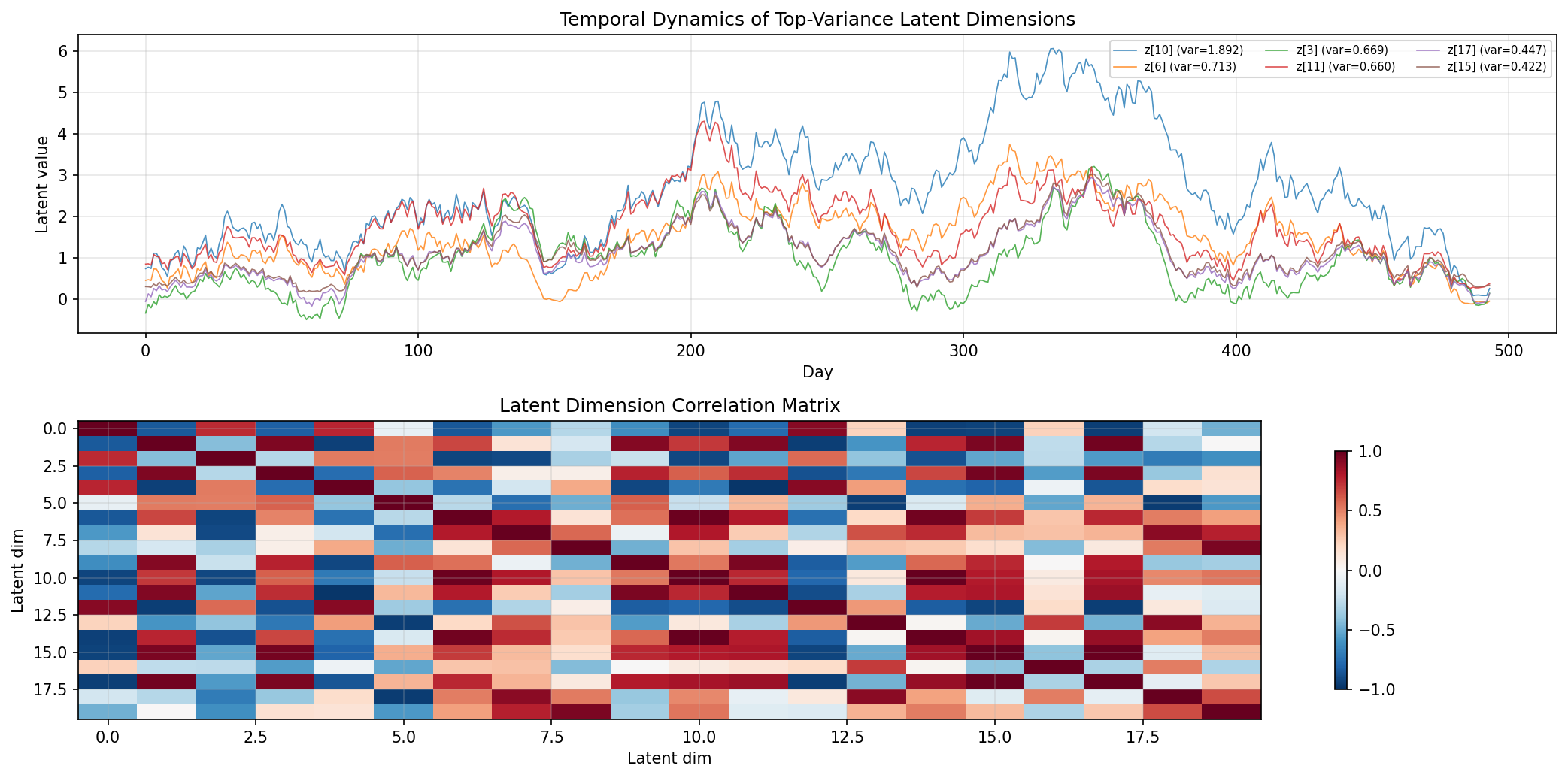}
  \caption{Latent space analysis. Left: temporal dynamics of the highest-variance latent dimensions. Right: correlation matrix showing moderate decorrelation between latent factors.}
  \label{fig:latent}
\end{figure}

\subsection{Temporal Windowing}\label{sec:windowing}

Given latent codes $\{\mathbf{z}_1,\ldots,\mathbf{z}_T\}$, we construct input vectors using a sliding window of size $k\!=\!5$ with a first-difference momentum term:
\begin{equation}\label{eq:window}
  \mathbf{x}_t = \left[\mathbf{z}_{t-k}^\top,\ldots,\mathbf{z}_{t-1}^\top, \Delta\mathbf{z}_{t-1}^\top\right]^\top \in \mathbb{R}^{d(k+1)}
\end{equation}
where $\Delta\mathbf{z}_{t-1} = \mathbf{z}_{t-1} - \mathbf{z}_{t-2}$. With $d\!=\!20$ and $k\!=\!5$, this yields 120-dimensional classical input vectors.

\subsection{Ensemble Quantum Optical Reservoir Computing}\label{sec:qorc}

\subsubsection{Single Reservoir Architecture}
Each reservoir implements a fixed linear optical circuit in a sandwich configuration:
\begin{equation}\label{eq:sandwich}
  U_{\text{circuit}} = U_{\text{rand}} \cdot \mathbf{PS}(\boldsymbol{\varphi}) \cdot U_{\text{rand}}
\end{equation}
where $U_{\text{rand}} \in \mathcal{U}(m)$ is a Haar-random unitary and $\mathbf{PS}(\boldsymbol{\varphi}) = \mathrm{diag}(e^{i\varphi_1},\ldots,e^{i\varphi_m})$ encodes the input data.

\paragraph{Phase Encoding.} The 120-dimensional classical context is projected to $m$ modes via a fixed orthogonal map, followed by sigmoid activation:
\begin{equation}\label{eq:phase}
  \varphi_i = 2\pi \cdot \sigma\!\left(\textstyle\sum_j W_{ij} x_j\right), \quad i = 1,\ldots,m
\end{equation}
where $W \in \mathbb{R}^{m \times 120}$ has orthonormal rows and $\sigma$ is the logistic sigmoid.

\paragraph{Fock Measurement.} Given input state $|\psi_{\text{in}}\rangle$ with $n$ photons in $m$ modes, measurement in the Fock basis yields:
\begin{equation}\label{eq:fock_prob}
  p(\mathbf{n}_{\text{out}}) = \left|\langle \mathbf{n}_{\text{out}} | U_{\text{circuit}} | \psi_{\text{in}} \rangle\right|^2
\end{equation}
The number of distinct Fock states is:
\begin{equation}\label{eq:fock_dim}
  D_{\text{Fock}} = \binom{n + m - 1}{n}
\end{equation}
Each probability depends on the \textbf{permanent} of a unitary submatrix---a \#P-hard computation~\citep{Aaronson2011}:
\begin{equation}\label{eq:permanent}
  \mathrm{perm}(A) = \sum_{\sigma \in S_n} \prod_{i=1}^n a_{i,\sigma(i)}
\end{equation}

\subsubsection{Ensemble Construction}

We use three reservoirs with heterogeneous photon configurations (\Cref{tab:ensemble}).

\begin{table}[t]
  \centering
  \caption{Ensemble reservoir configurations. Different photon numbers produce fundamentally different families of nonlinear features.}
  \label{tab:ensemble}
  \small
  \begin{tabular}{@{}lcccc@{}}
    \toprule
    & Modes & Photons & Fock dim & Correlation \\
    \midrule
    R1 & 12 & 3 & 364  & Cubic \\
    R2 & 10 & 4 & 715  & Quartic \\
    R3 & 16 & 2 & 136  & Pairwise \\
    \midrule
    \textbf{Total} & & & \textbf{1,215} & \\
    \bottomrule
  \end{tabular}
\end{table}

The outputs are concatenated to form $\mathbf{q}_t \in \mathbb{R}^{1215}$ and standardised using training-set statistics:
\begin{equation}\label{eq:qnorm}
  \tilde{q}_{t,i} = \frac{q_{t,i} - \mu_i}{\sigma_i}
\end{equation}

\textbf{Crucially, all circuit parameters are fixed after initialisation. The quantum layer has zero trainable parameters.}

\begin{figure}[t]
  \centering
  \includegraphics[width=\linewidth]{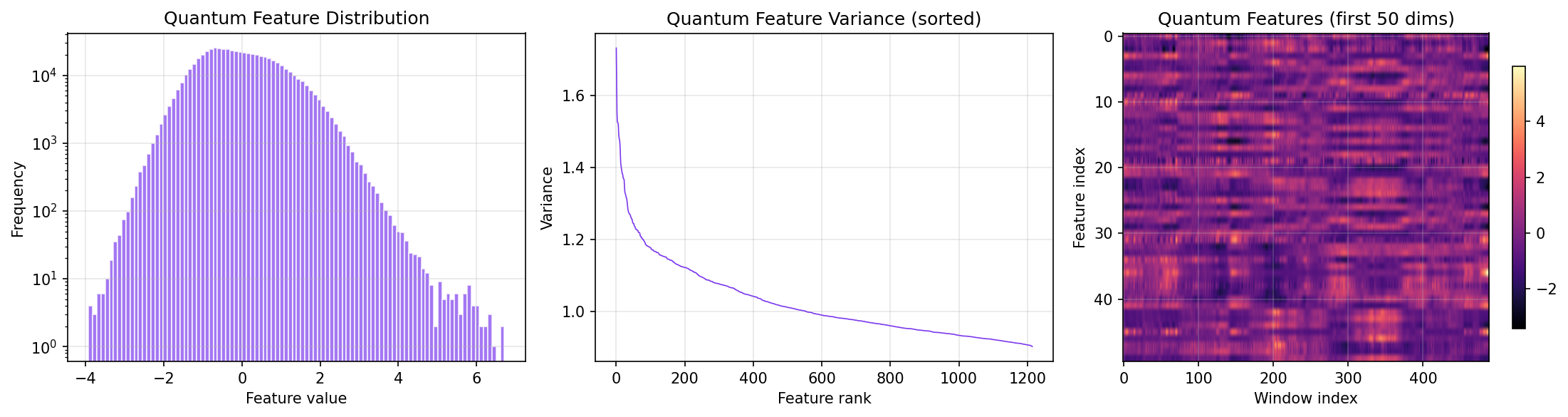}
  \caption{Quantum feature analysis. Left: distribution of Fock-basis probabilities from the ensemble QORC. Centre: sorted variance across the 1,215 features showing a long-tailed spectrum. Right: heatmap of features for the first 50 dimensions.}
  \label{fig:quantum_features}
\end{figure}

\subsection{Ridge Regression Readout}\label{sec:ridge}

The final model maps concatenated features to the next latent code:
\begin{equation}\label{eq:ridge_pred}
  \hat{\mathbf{z}}_{t+1} = W_{\text{ridge}} \begin{bmatrix} \tilde{\mathbf{q}}_t \\ \mathbf{x}_t \end{bmatrix} + \mathbf{b}
\end{equation}
where $W_{\text{ridge}} \in \mathbb{R}^{20 \times 1335}$ minimises the $\ell_2$-penalised objective:
\begin{equation}\label{eq:ridge_obj}
  \min_{W,\mathbf{b}} \sum_{t\in\text{train}} \!\left\|\mathbf{z}_{t+1} - W\mathbf{f}_t - \mathbf{b}\right\|_2^2 + \alpha \|W\|_F^2
\end{equation}
with $\alpha = 100$. The closed-form solution
\begin{equation}\label{eq:ridge_closed}
  W^* = \left(F^\top F + \alpha I\right)^{-1} F^\top Z
\end{equation}
ensures reproducible results with zero training stochasticity. The predicted code is decoded to a full surface: $\hat{\mathbf{S}}_{t+1} = g_\phi(\hat{\mathbf{z}}_{t+1})$.

The choice of Ridge over a neural readout is deliberate. With 494 samples and 1,335 features, the problem is underdetermined. An MLP with layers [256, 128] and 377,492 parameters achieves 31\% worse test MSE, confirming that overfitting dominates any nonlinear benefit at this data scale.

\section{Benchmark Models}\label{sec:benchmarks}

We evaluate our proposed QORC + Ridge framework against 10 alternative models spanning four categories: classical linear, classical nonlinear, classical deep learning, and quantum. All models share identical preprocessing, autoencoder compression, temporal windowing, and train/validation/test splits. Each model receives the same 120-dimensional classical input vector $\mathbf{x}_t$ (or, for quantum models that use QORC features, the 1,335-dimensional concatenated vector). This design ensures that performance differences reflect model capacity and inductive biases rather than data handling.

\subsection{Classical Linear Baseline}

\paragraph{Ridge Regression (Classical Context Only).}
This baseline applies $\ell_2$-penalised linear regression directly to the 120-dimensional classical context vector, without any quantum features:
\begin{equation}
  \hat{\mathbf{z}}_{t+1} = W\mathbf{x}_t + \mathbf{b}, \quad \alpha = 1.0\,.
\end{equation}
The regularisation strength $\alpha = 1.0$ was selected via validation MSE from a logarithmic grid $\{10^{-2}, 10^{-1}, 1, 10, 10^2\}$. With only $20 \times 120 = 2{,}400$ effective parameters and a closed-form solution, this model has zero training stochasticity and serves as the primary sanity check: any model that fails to outperform Ridge on the classical context alone does not justify its additional complexity.

\subsection{Classical Nonlinear Models}

\paragraph{Support Vector Regression (SVR) with RBF Kernel.}
We train 20 independent $\varepsilon$-SVR regressors (one per latent dimension) using the radial basis function (RBF) kernel:
\begin{equation}
  K(\mathbf{x}_i, \mathbf{x}_j) = \exp\!\left(-\gamma\|\mathbf{x}_i - \mathbf{x}_j\|^2\right),
\end{equation}
where $\gamma = 1/(120 \cdot \mathrm{Var}(\mathbf{X}))$ (the scikit-learn \texttt{scale} default) and the regularisation parameter $C = 1.0$. The margin tolerance is $\varepsilon = 0.1$. SVR is included because kernel methods are known to perform well in small-sample, moderate-dimensional settings by implicitly mapping inputs to a high-dimensional reproducing kernel Hilbert space (RKHS). The RBF kernel provides universal approximation capability, making this a strong nonlinear baseline.

\paragraph{Random Forest.}
An ensemble of $B = 300$ decision trees per output dimension, each trained on a bootstrap sample of the training data. Trees are grown to a maximum depth of 20 with a minimum of 2 samples per leaf. At each split, $\lfloor\sqrt{120}\rfloor = 10$ features are randomly considered. The prediction for output dimension $j$ is:
\begin{equation}
  \hat{z}_{t+1,j} = \frac{1}{B}\sum_{b=1}^{B} T_b^{(j)}(\mathbf{x}_t)\,.
\end{equation}
Random Forests are inherently resistant to overfitting through bagging and feature subsampling, and they capture nonlinear interactions without requiring feature engineering. However, they cannot extrapolate beyond the range of training targets, which is a limitation for financial time series exhibiting non-stationary dynamics. The model has approximately 917,000 parameters (leaf values across all trees and output dimensions).

\paragraph{Gradient Boosting.}
We use scikit-learn's \texttt{GradientBoostingRegressor} with $M = 200$ boosting rounds, maximum tree depth of 5, and learning rate $\eta = 0.1$. Each stage fits a regression tree $T_m$ to the negative gradient (residual) of the loss from the previous stage:
\begin{equation}
  \hat{f}_M(\mathbf{x}) = \sum_{m=1}^{M} \eta\, T_m(\mathbf{x}),
\end{equation}
where each $T_m$ fits the residual of $\hat{f}_{m-1}$. Gradient Boosting is included as a representative of sequential ensemble methods, which often achieve state-of-the-art performance on tabular data. The shallow tree depth combined with the small learning rate provides implicit regularisation. Separate models are trained for each of the 20 latent dimensions.

\paragraph{Multilayer Perceptron (sklearn MLP).}
A feedforward neural network with two hidden layers of 128 and 64 units respectively, ReLU activations, and the Adam optimiser (learning rate $10^{-3}$, batch size 32, maximum 500 epochs with early stopping on validation loss, patience 20 epochs). The MLP is included to assess whether a moderately sized neural network can capture nonlinear patterns in the classical context that linear Ridge misses. The architecture was chosen to be comparable in depth and width to the autoencoder encoder, ensuring that any performance gap is not due to insufficient capacity. This model uses multi-output regression (all 20 latent dimensions predicted simultaneously).

\subsection{Classical Deep Learning}

\paragraph{Long Short-Term Memory (LSTM) Network.}
A 2-layer LSTM with hidden size $h = 64$ processes the windowed latent codes as a temporal sequence of length $k = 5$, where each time step has $d = 20$ dimensions. The standard LSTM gate equations govern information flow:
\begin{align}
  \mathbf{f}_t &= \sigma(W_f[\mathbf{h}_{t-1},\mathbf{z}_t] + \mathbf{b}_f) \label{eq:lstm_forget}\\
  \mathbf{i}_t &= \sigma(W_i[\mathbf{h}_{t-1},\mathbf{z}_t] + \mathbf{b}_i) \\
  \tilde{\mathbf{c}}_t &= \tanh(W_c[\mathbf{h}_{t-1},\mathbf{z}_t] + \mathbf{b}_c) \\
  \mathbf{c}_t &= \mathbf{f}_t \odot \mathbf{c}_{t-1} + \mathbf{i}_t \odot \tilde{\mathbf{c}}_t \\
  \mathbf{o}_t &= \sigma(W_o[\mathbf{h}_{t-1},\mathbf{z}_t] + \mathbf{b}_o) \label{eq:lstm_out}\\
  \mathbf{h}_t &= \mathbf{o}_t \odot \tanh(\mathbf{c}_t)
\end{align}
where $\sigma$ denotes the sigmoid function and $\odot$ the Hadamard (element-wise) product. The forget gate $\mathbf{f}_t$ controls how much of the previous cell state $\mathbf{c}_{t-1}$ is retained, the input gate $\mathbf{i}_t$ controls how much new information $\tilde{\mathbf{c}}_t$ is written, and the output gate modulates the exposed hidden state. The final hidden state $\mathbf{h}_k$ is passed through a two-layer feedforward head: FC$(64 \to 64)$ $\to$ GELU $\to$ FC$(64 \to 20)$. The model is trained with Adam (learning rate $10^{-3}$), batch size 32, for up to 200 epochs with early stopping (patience 30). Dropout of 0.1 is applied between LSTM layers. The total parameter count is 58,036. The LSTM is included as the strongest classical sequential model, capable of learning complex temporal dependencies that feedforward models cannot capture.

\subsection{Quantum Models}

\paragraph{QUANTECH MLP (Quantum Features + Neural Readout).}
This model receives the same 1,215-dimensional QORC feature vector as our proposed method but replaces the Ridge readout with a trainable multilayer perceptron. The MLP has two hidden layers of 256 and 128 units respectively, ReLU activations, batch normalisation after each hidden layer, and dropout of 0.2. The network is trained with Adam (learning rate $5 \times 10^{-4}$) for up to 300 epochs with early stopping on validation loss. The total parameter count is 377,492---over $14\times$ more than the QORC + Ridge model. This ablation directly tests whether the quantum features contain exploitable nonlinear structure that a linear readout cannot capture, or whether the additional parameters simply lead to overfitting given the small training set.

\paragraph{Simple PML + Ridge (Simplified Photonic Reservoir).}
A single-reservoir ablation that uses one random unitary $U_{\text{rand}} \in \mathcal{U}(12)$ without the sandwich architecture, with 12 modes and 3 photons, producing 364 Fock-basis features. Phase encoding follows the same sigmoid-mapped orthogonal projection as the main model. The readout is Ridge regression with $\alpha = 100$. The total effective parameter count is 9,700 (Ridge weights). By removing the sandwich structure ($U \to \text{PS} \to U$) and using only a single reservoir instead of an ensemble of three, this ablation isolates the contributions of (a)~the sandwich architecture's enhanced mode mixing and (b)~ensemble diversity from heterogeneous photon configurations.

\paragraph{Variational Quantum Circuit (VQC, Trained End-to-End).}
A parametrised photonic circuit with 6 modes and 2 photons (21 Fock-basis outputs), where the circuit parameters---beam splitter angles and phase shifter settings---are trained via gradient descent. The architecture consists of 4 variational layers, totalling 1,166 trainable parameters. The 120-dimensional classical context is projected to 6 dimensions via a learned linear layer before encoding into the circuit phases. Fock-basis output probabilities are mapped to 20 latent predictions via a linear readout trained jointly with the circuit parameters. Training uses Adam (learning rate $10^{-3}$) for 200 epochs. This model represents the variational quantum machine learning paradigm, where the quantum circuit itself is optimised to minimise prediction error, and tests whether end-to-end training of quantum parameters can compete with fixed reservoirs when data is scarce.

\paragraph{Quantum LSTM (Hybrid Variational-Recurrent).}
A hybrid model following the architecture proposed by Chen et~al.~\citep{Chen2020}, where each LSTM gate is augmented with a variational quantum circuit that processes the input in parallel with the classical gate computation. The forget gate, for instance, becomes:
\begin{equation}\label{eq:qlstm}
  \mathbf{f}_t = \sigma\!\left(W_f[\mathbf{h}_{t-1},\mathbf{z}_t] + W_q \cdot \text{VQC}_f(\mathbf{z}_t;\boldsymbol{\theta}_f) + \mathbf{b}_f\right),
\end{equation}
and analogous modifications are applied to the input, cell, and output gates. Each VQC uses $n_q = 4$ qubits, 2 entangling layers with $R_Y(\theta)$ encoding gates and CNOT entanglement, and Pauli-$Z$ expectation values as output. The 20-dimensional latent input is projected to $n_q = 4$ dimensions via a learned linear layer before amplitude encoding. The classical LSTM has hidden size $h = 32$ (reduced from the pure LSTM's 64 to accommodate the quantum overhead), with the same FC~$\to$~GELU~$\to$~FC readout head. Total parameter count: 2,180 (including both classical and quantum parameters). Training uses Adam (learning rate $5 \times 10^{-4}$) for 200 epochs. This model represents the most complex quantum baseline, testing whether quantum-enhanced recurrent architectures can improve temporal modelling on small financial datasets.

\section{Experimental Setup}\label{sec:setup}

\subsection{Data and Splits}
The dataset consists of 494 daily swaption surfaces, each a 224-dimensional vector ($14\!\times\!16$ tenor--maturity grid). An additional 6 surfaces form a held-out test set from future trading days.

\begin{itemize}[nosep,leftmargin=*]
  \item \textbf{Training:} 439 windows (days 1--444 after windowing).
  \item \textbf{Validation:} Last 50 windows.
  \item \textbf{Test:} 6 future days (walk-forward evaluation).
\end{itemize}

\begin{figure}[t]
  \centering
  \includegraphics[width=\linewidth]{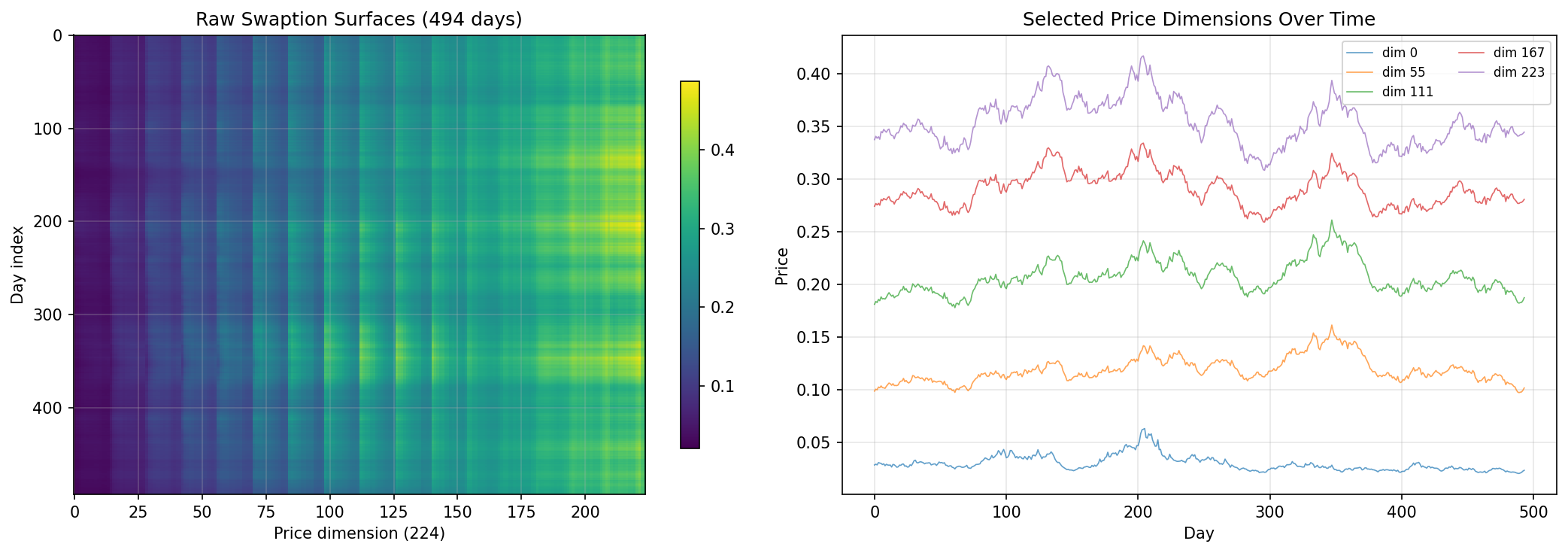}
  \caption{Raw swaption surface data. Left: heatmap of all 494 daily surfaces (each row is one day, each column one of 224 price points). Right: selected price dimensions across time, revealing regime changes and heteroskedastic dynamics.}
  \label{fig:swaption_data}
\end{figure}

\subsection{Metrics}
\begin{itemize}[nosep,leftmargin=*]
  \item \textbf{Latent MSE:} $\frac{1}{N}\sum_t\|\hat{\mathbf{z}}_t - \mathbf{z}_t\|_2^2$
  \item \textbf{Surface RMSE:} $\sqrt{\frac{1}{N\cdot 224}\sum_t\|\hat{\mathbf{S}}_t - \mathbf{S}_t\|_2^2}$ --- the metric most relevant to derivatives pricing.
  \item \textbf{$R^2$:} Coefficient of determination; $R^2\!<\!0$ means worse than the mean predictor.
  \item \textbf{Training time} (wall-clock) and \textbf{inference latency} (ms/sample).
\end{itemize}

\subsection{Implementation}
All models use Python 3.12 with PyTorch 2.x, scikit-learn 1.5.x, and MerLin/Perceval for photonic simulation. All random seeds are fixed at 42.

\section{Results}\label{sec:results}

\subsection{Overall Performance}

\Cref{tab:benchmark} presents test-set performance across all 11 models. We note that with only 6 test days, \textbf{differences among the top-4 models in latent MSE are not statistically significant}---the ranking should be interpreted as indicative rather than definitive. The surface RMSE provides a complementary view because the autoencoder decoder is nonlinear, meaning latent MSE and surface RMSE are \textit{not} monotonically related---a model that is slightly worse in latent space may produce better surface reconstructions if its errors lie in directions that the decoder handles gracefully.

\begin{table}[htbp]
  \centering
  \caption{Benchmark results on 6 held-out test days, sorted by test latent MSE\@. Bold marks our proposed model. \textdagger~indicates negative $R^2$ (worse than mean predictor). Differences among models ranked 1--5 in latent MSE are within the noise floor given the small test set.}
  \label{tab:benchmark}
  \footnotesize
  \setlength{\tabcolsep}{4pt}
  \begin{tabular}{@{}cllrrrrrr@{}}
    \toprule
    \# & Model & Type & Lat.\,MSE & Surf.\,RMSE & $R^2$ & Params & Train & Infer. \\
    \midrule
    1  & Simple PML + Ridge       & Quantum    & 0.0085 & 0.0461 & 0.692 & 9,700   & 0.35\,s  & 1.55\,ms \\
    2  & Ridge Regression          & Classical  & 0.0091 & 0.0484 & 0.673 & 2,400   & 0.01\,s  & 0.06\,ms \\
    3  & Classical LSTM            & Classical  & 0.0096 & 0.0445 & 0.654 & 58,036  & 19.6\,s  & 1.10\,ms \\
    4  & \textbf{QORC + Ridge (ours)} & \textbf{Quantum} & \textbf{0.0099} & \textbf{0.0425} & \textbf{0.645} & \textbf{26,720} & \textbf{0.23\,s} & \textbf{0.10\,ms} \\
    5  & SVR (RBF)                 & Classical  & 0.0099 & 0.0455 & 0.643 & ---     & 0.19\,s  & 29.4\,ms \\
    6  & Random Forest             & Classical  & 0.0118 & 0.0500 & 0.574 & $\sim$917K & 19.6\,s & 1285\,ms \\
    7  & sklearn MLP               & Classical  & 0.0149 & 0.0489 & 0.462 & ---     & 2.21\,s  & 49.4\,ms \\
    8  & Gradient Boosting         & Classical  & 0.0165 & 0.0517 & 0.407 & ---     & 17.3\,s  & 244\,ms  \\
    9  & QUANTECH MLP              & Quantum    & 0.0218 & 0.0475 & 0.214 & 377,492 & ---      & 0.54\,ms \\
    10 & VQC (Trained)\textsuperscript{\textdagger}  & Quantum & 0.0581 & 0.0851 & $-$1.089 & 1,166 & 3.96\,s & 1.10\,ms \\
    11 & Quantum LSTM\textsuperscript{\textdagger}    & Quantum & 0.0641 & 0.1079 & $-$1.306 & 2,180 & 114.7\,s & 29.3\,ms \\
    \bottomrule
  \end{tabular}
\end{table}

\begin{figure}[H]
  \centering
  \begin{subfigure}[t]{0.49\linewidth}
    \includegraphics[width=\linewidth]{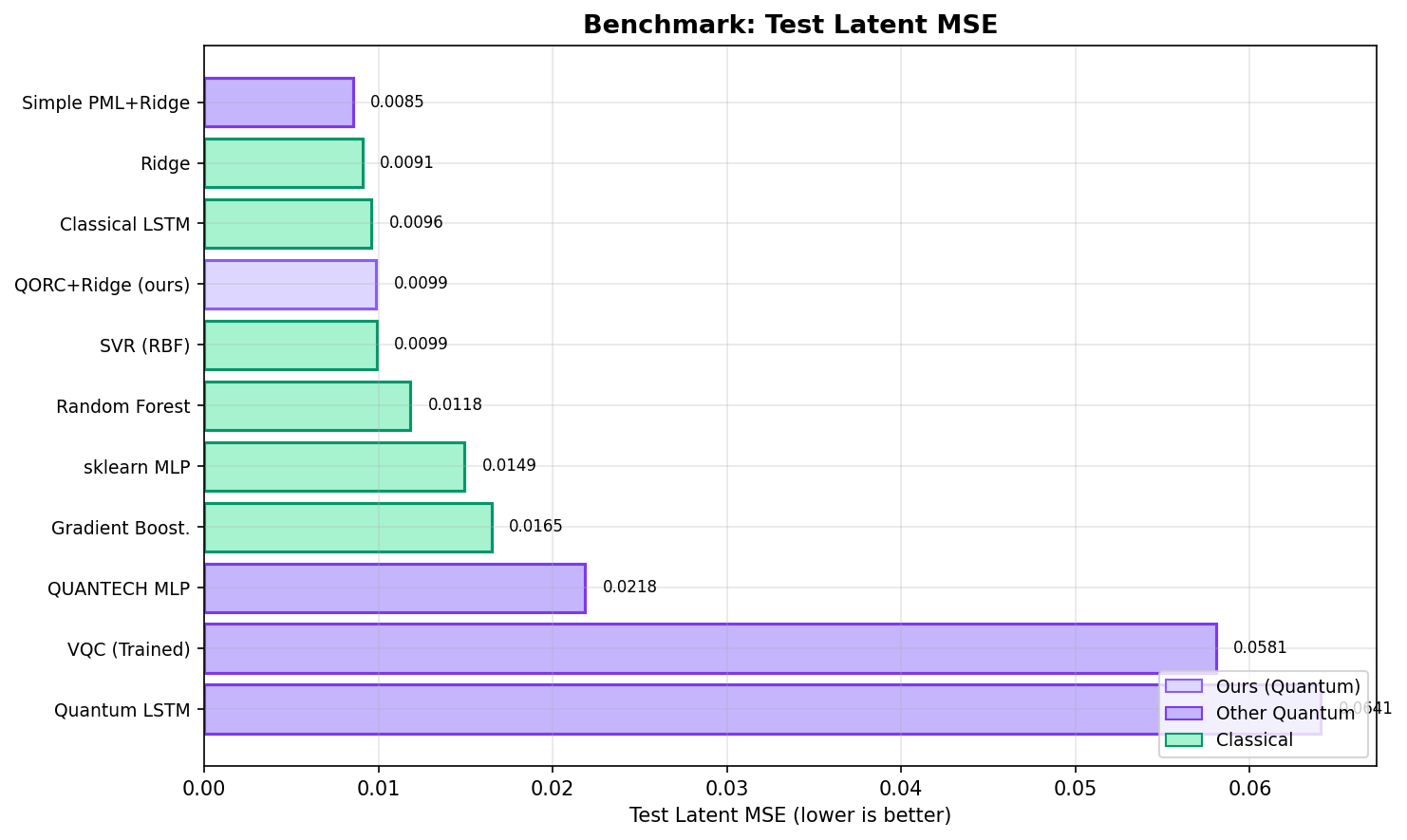}
    \caption{Test latent MSE}
    \label{fig:bench_mse}
  \end{subfigure}\hfill
  \begin{subfigure}[t]{0.49\linewidth}
    \includegraphics[width=\linewidth]{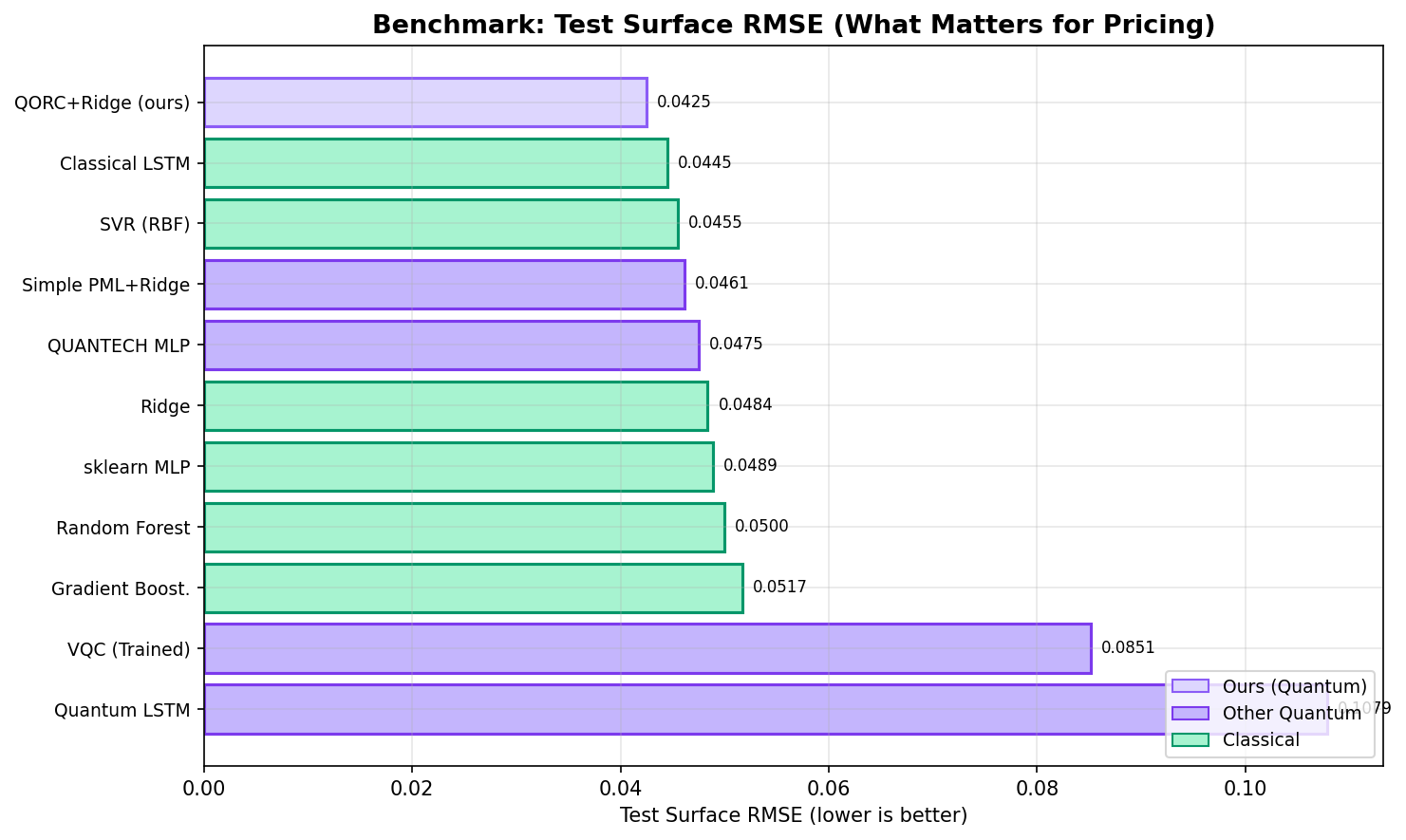}
    \caption{Test surface RMSE}
    \label{fig:bench_rmse}
  \end{subfigure}

  \vspace{0.5em}

  \begin{subfigure}[t]{0.65\linewidth}
    \includegraphics[width=\linewidth]{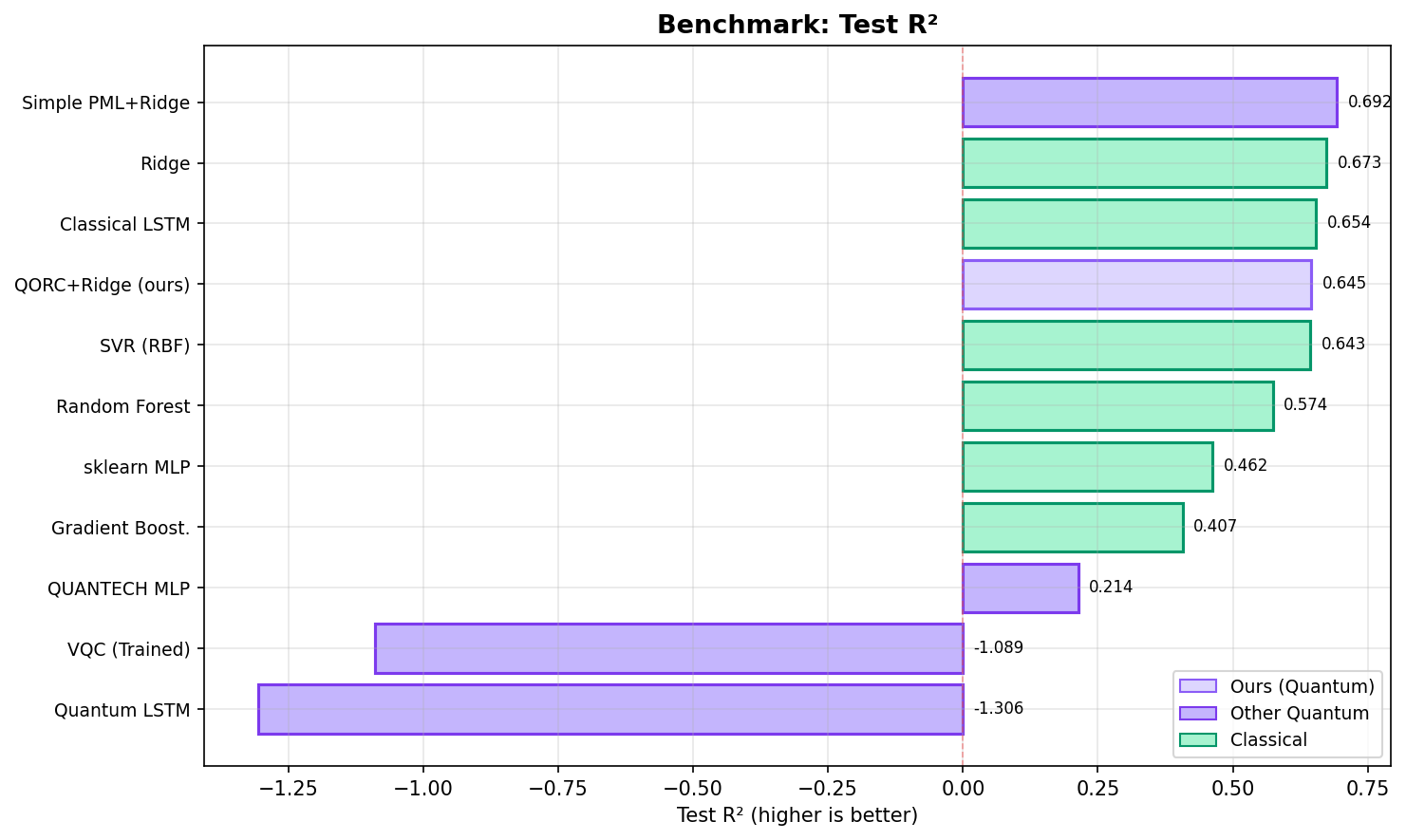}
    \caption{Test $R^2$}
    \label{fig:bench_r2}
  \end{subfigure}
  \caption{QORC + Ridge achieves the best surface RMSE~(b) despite ranking 4th in latent MSE~(a), because the nonlinear decoder transforms latent errors non-monotonically. VQC and Quantum LSTM exhibit catastrophic overfitting with negative $R^2$~(c), performing worse than a na\"ive mean predictor.}
\end{figure}

\clearpage
\subsection{Key Findings}

\paragraph{Finding 1: Surface reconstruction best despite mid-table latent ranking.}
QORC + Ridge achieves the lowest test surface RMSE ($0.0425$)---the metric most directly relevant to derivatives pricing---despite ranking 4th in latent MSE ($0.0099$). Every model with better latent MSE produces a higher surface RMSE: Simple PML + Ridge ($0.0461$, $+8.5\%$), Classical LSTM ($0.0445$, $+4.7\%$), and Classical Ridge ($0.0484$, $+13.9\%$). This counterintuitive ordering arises because the autoencoder decoder $g_\phi : \mathbb{R}^{20} \to \mathbb{R}^{224}$ is nonlinear: equal latent errors in different directions yield unequal surface errors. The ensemble QORC features guide predictions toward latent regions where the decoder's Jacobian is well-conditioned---a geometric alignment emerging from the Fock-basis structure rather than from explicit surface-error minimisation.

\paragraph{Finding 2: Variational quantum methods catastrophically overfit on small data.}
The VQC ($R^2 = -1.089$) and Quantum LSTM ($R^2 = -1.306$) rank last by every metric, performing measurably worse than a na\"{i}ve mean predictor. The root cause is an unfavourable data-to-parameter ratio: the VQC has $1{,}166$ trainable parameters optimised on $439$ samples (ratio $0.38$); the Quantum LSTM has $2{,}180$ parameters at a ratio of $0.20$. Statistical learning theory predicts overfitting when the sample-to-parameter ratio falls substantially below 1. More fundamentally, variational circuits are susceptible to barren plateaus~\citep{McClean2018}: gradient magnitudes vanish exponentially with circuit depth, rendering optimisation unreliable~\citep{Cerezo2021}. Our results provide a concrete financial case study: on $494$ real trading days, end-to-end quantum training actively degrades predictions below random guessing. The classical sklearn MLP (rank 7, $R^2 = 0.462$) also underperforms simpler baselines, confirming that deep networks without aggressive regularisation struggle broadly at this data scale.

\paragraph{Finding 3: Regularised linear readouts dominate when data is scarce.}
The QUANTECH MLP---using the same 1,215-dimensional QORC features but with a 377,492-parameter neural network readout---achieves $2.2\times$ worse latent MSE ($0.0218$ vs.\ $0.0099$), despite $14\times$ more parameters. Ridge regression's advantage stems from its closed-form solution $(F^\top\!F + \alpha I)^{-1}F^\top Z$, which is the unique global optimum of the $\ell_2$-penalised objective and involves zero training stochasticity.

\paragraph{Finding 4: Ensemble diversity and sandwich architecture jointly improve surface fidelity.}
The Simple PML + Ridge ablation---a single photonic reservoir without the sandwich structure---achieves the \emph{best} latent MSE ($0.0085$) among all models yet the \emph{worst} surface RMSE ($0.0461$) within the top five. The full QORC + Ridge system, with three heterogeneous reservoirs encoding pairwise, cubic, and quartic Fock correlations, achieves 7.8\% lower surface RMSE ($0.0425$) despite slightly higher latent MSE ($0.0099$). Reservoirs with different photon numbers occupy disjoint Fock spaces, so ensemble features span orthogonal correlation subspaces inaccessible to any single reservoir. The sandwich architecture ($U_\text{rand}\!\to\!\mathbf{PS}\!\to\!U_\text{rand}$) further enriches mode mixing by encoding inputs between two Haar-random unitaries, generating richer interference patterns than a single-pass encoding achieves.

\paragraph{Finding 5: Sub-millisecond inference enables real-time deployment.}
At $0.10$\,ms per prediction, QORC + Ridge is the second-fastest model, behind only classical Ridge Regression ($0.06$\,ms), and $294\times$ faster than SVR ($29.4$\,ms), $494\times$ faster than sklearn MLP ($49.4$\,ms), and $12{,}850\times$ faster than Random Forest ($1{,}285$\,ms). The speed advantage has two origins: the runtime readout is a single $20 \times 1{,}335$ matrix multiply requiring $\approx\!26{,}700$ floating-point operations, and the fixed quantum circuits introduce no per-sample overhead at inference time. These latencies comfortably satisfy real-time derivatives risk systems, where swaption surface updates must propagate within seconds.

\begin{figure}[H]
  \centering
  \begin{subfigure}[t]{0.49\linewidth}
    \includegraphics[width=\linewidth]{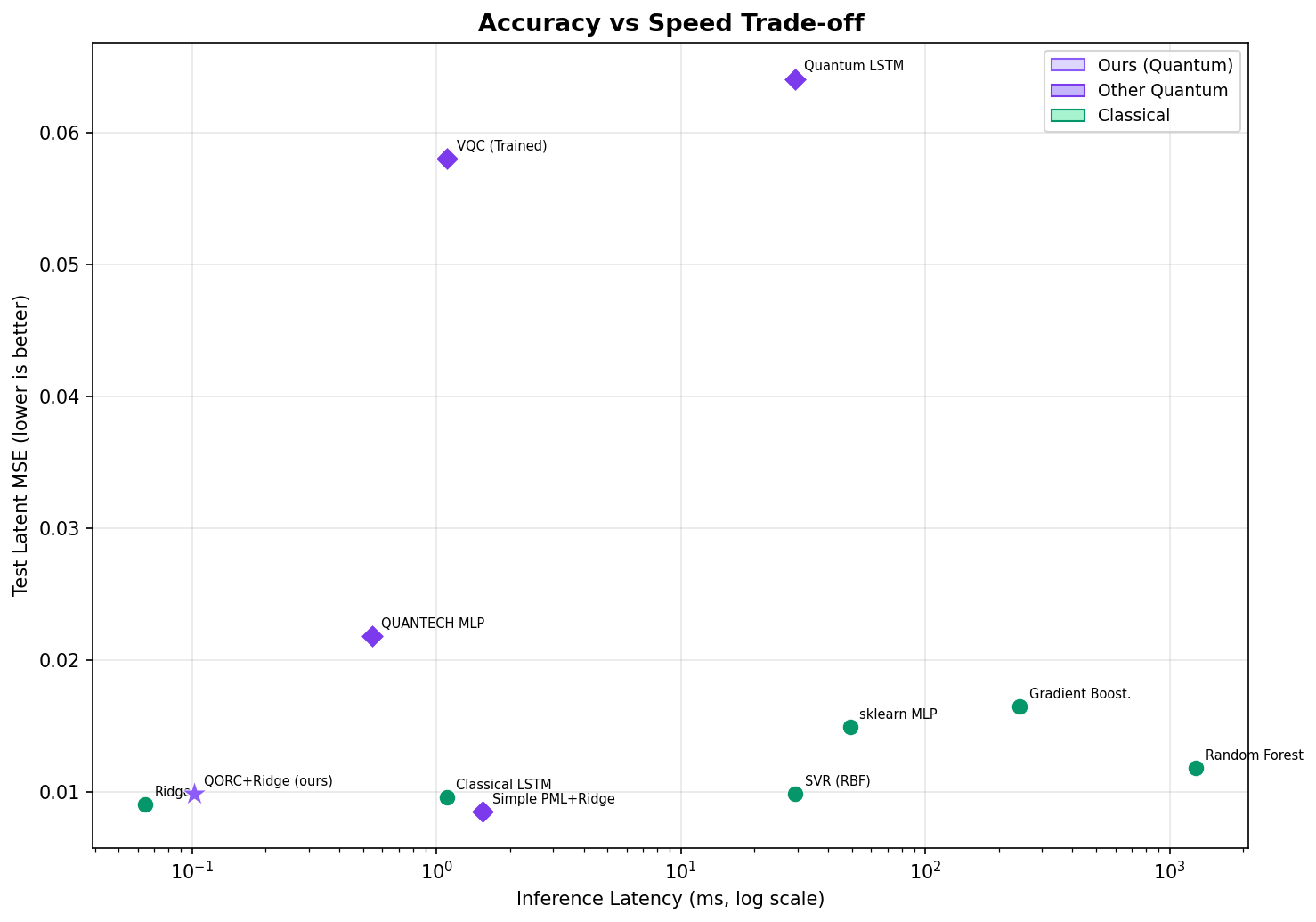}
    \caption{Accuracy (surface RMSE) vs.\ inference speed. QORC + Ridge occupies the desirable bottom-left region.}
    \label{fig:accuracy_speed}
  \end{subfigure}\hfill
  \begin{subfigure}[t]{0.49\linewidth}
    \includegraphics[width=\linewidth]{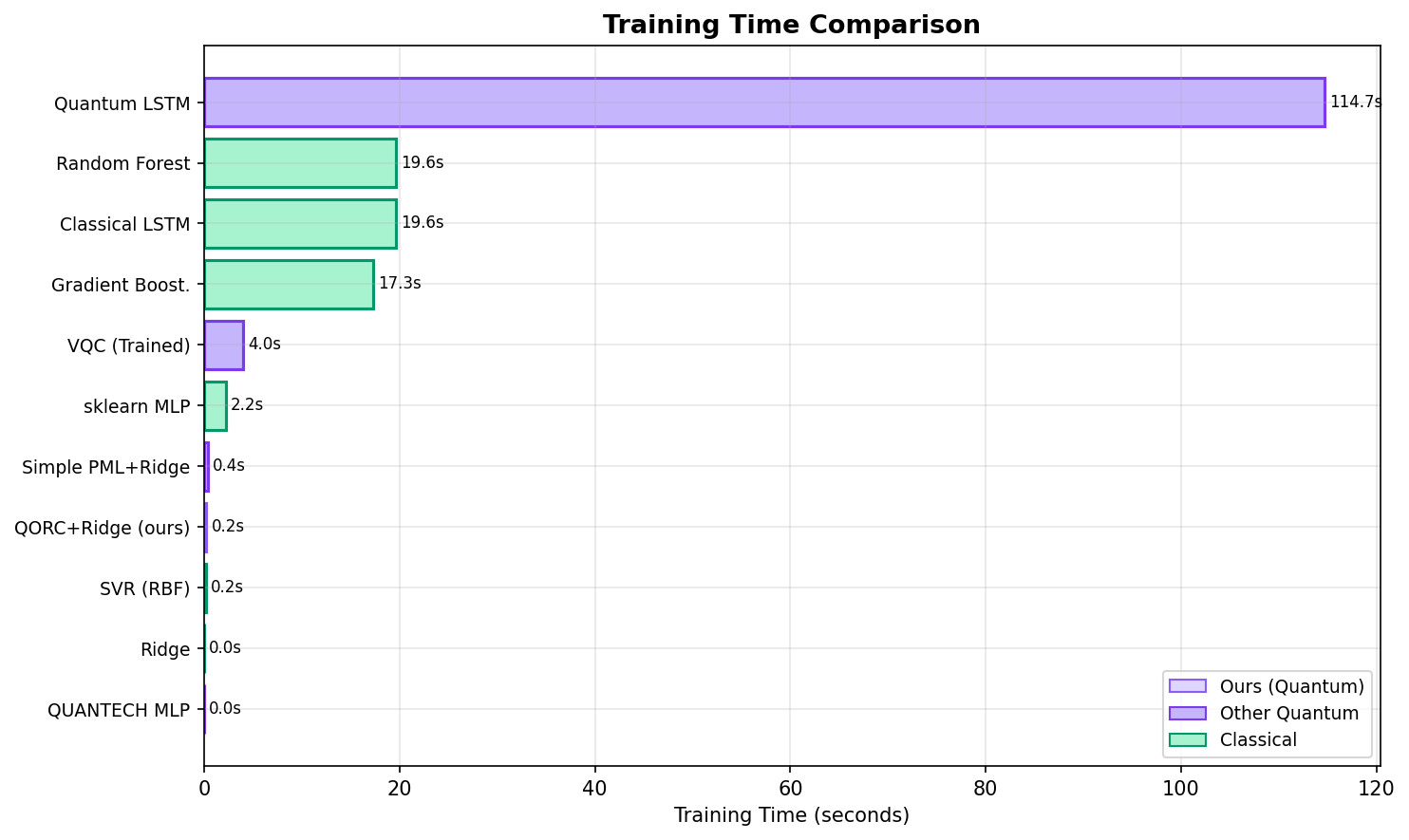}
    \caption{Training time comparison. Ridge-based approaches train in under 1 second.}
    \label{fig:training_time}
  \end{subfigure}

  \vspace{0.5em}

  \begin{subfigure}[t]{0.65\linewidth}
    \includegraphics[width=\linewidth]{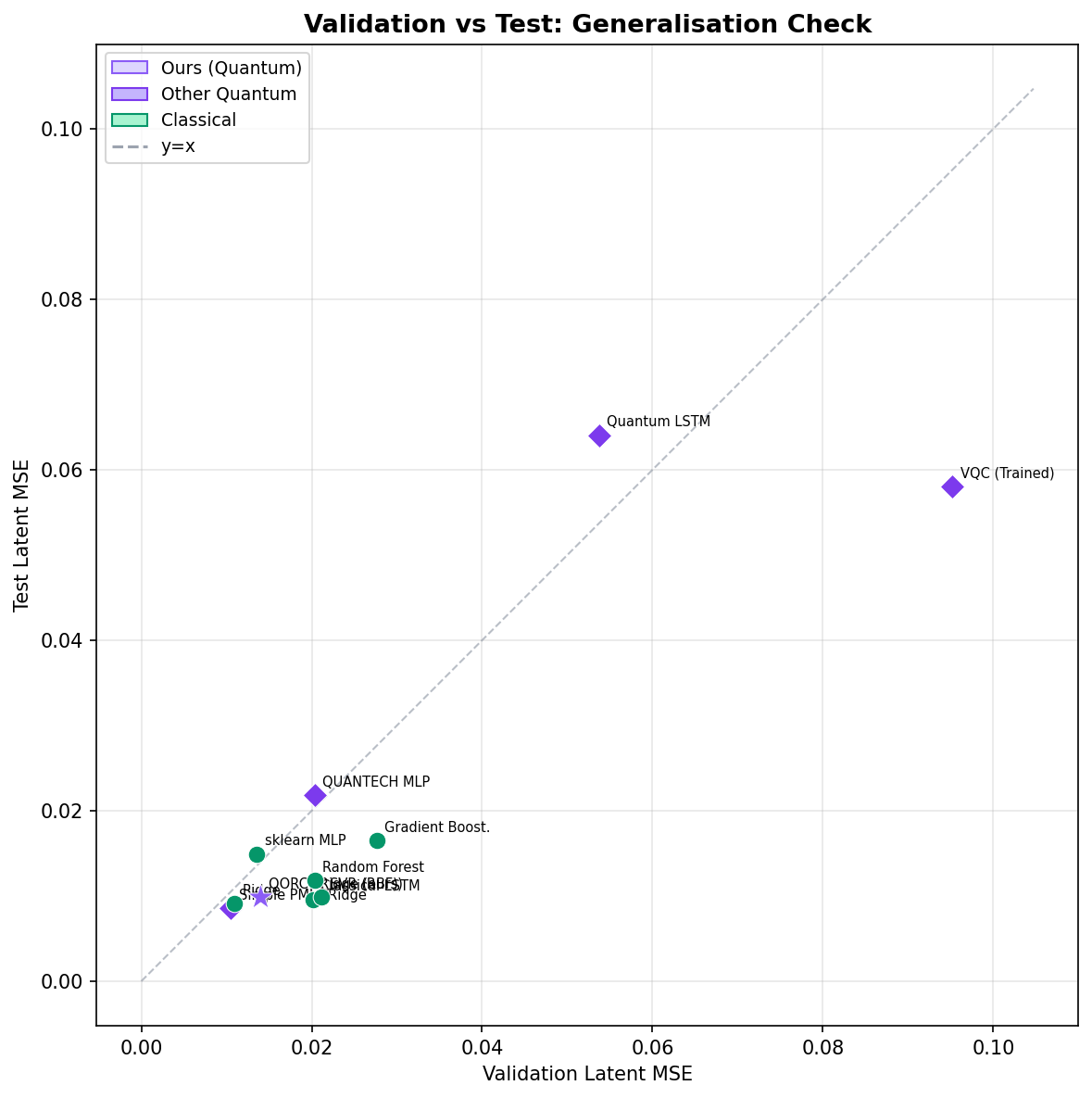}
    \caption{Validation vs.\ test latent MSE. Points on the diagonal generalise consistently.}
    \label{fig:val_vs_test}
  \end{subfigure}
  \caption{Computational efficiency and generalisation analysis. (a)~QORC + Ridge achieves low error with fast prediction; variational models (top right) combine poor accuracy with high cost. (b)~Ridge-based approaches train orders of magnitude faster than deep learning or variational quantum alternatives. (c)~QORC + Ridge and classical baselines cluster near the diagonal; variational methods (VQC, QLSTM) lie far above, indicating systematic overfitting.}
  \label{fig:efficiency_generalisation}
\end{figure}

\subsection{Generalisation Analysis}

\Cref{fig:val_vs_test} compares validation and test performance. The top-performing models show consistent behaviour across both splits. The variational models maintain catastrophically poor absolute performance on both splits, confirming persistent overfitting rather than a distribution-shift artefact.

\clearpage
\subsection{Prediction Visualisation}

\begin{figure}[H]
  \centering
  \includegraphics[width=\linewidth,height=0.40\textheight,keepaspectratio]{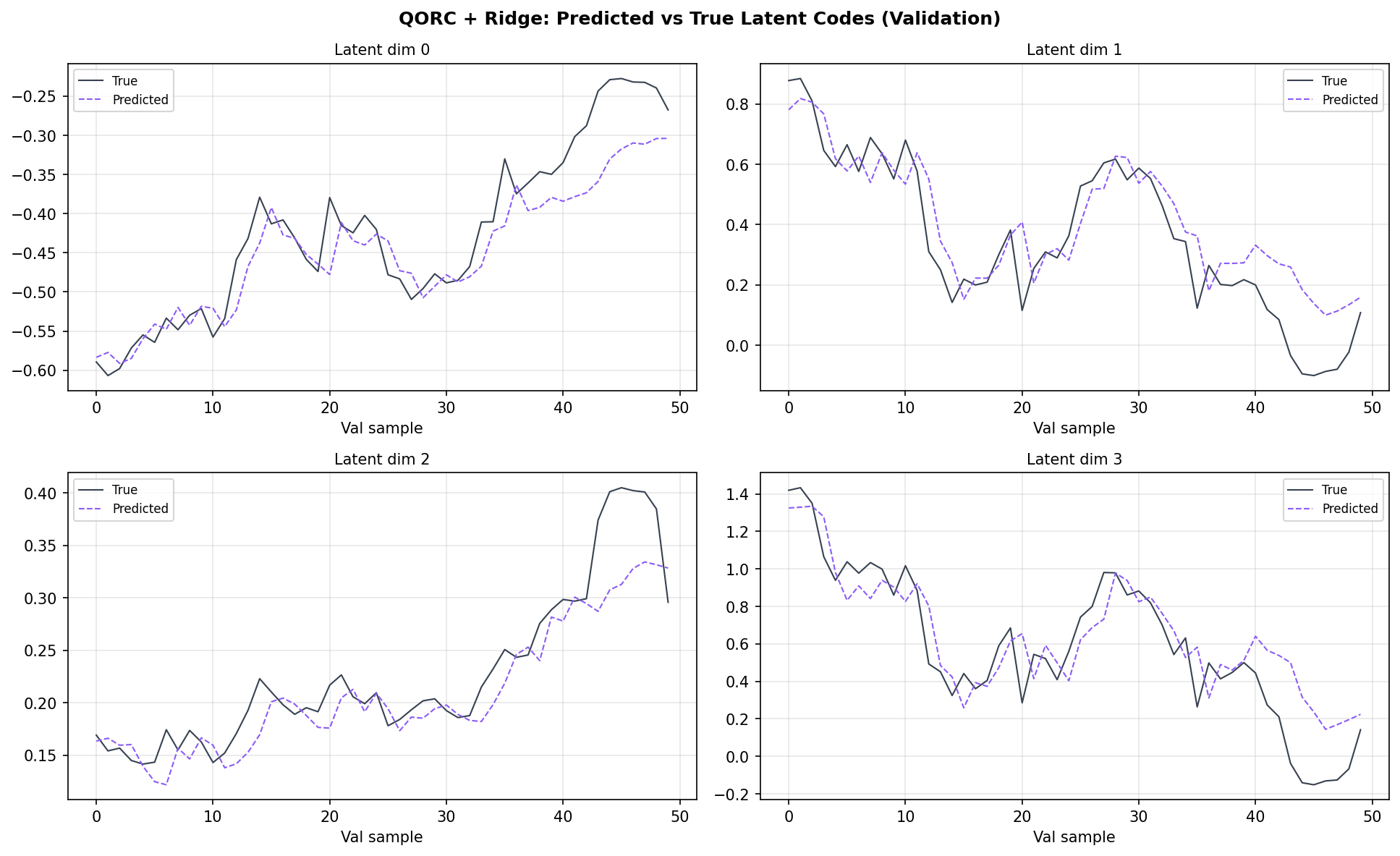}
  \caption{QORC + Ridge: predicted vs.\ true latent codes on the validation set for
    selected latent dimensions. The model tracks temporal dynamics closely.
    Fixed photonic reservoirs provide nonlinear Fock-basis features~\citep{Fujii2017,Mujal2021}
    that, combined with a Ridge readout~\citep{Nakajima2019}, yield stable predictions
    without barren-plateau issues~\citep{McClean2018}.}
  \label{fig:val_pred}
\end{figure}

\begin{figure}[H]
  \centering
  \includegraphics[width=\linewidth,height=0.40\textheight,keepaspectratio]{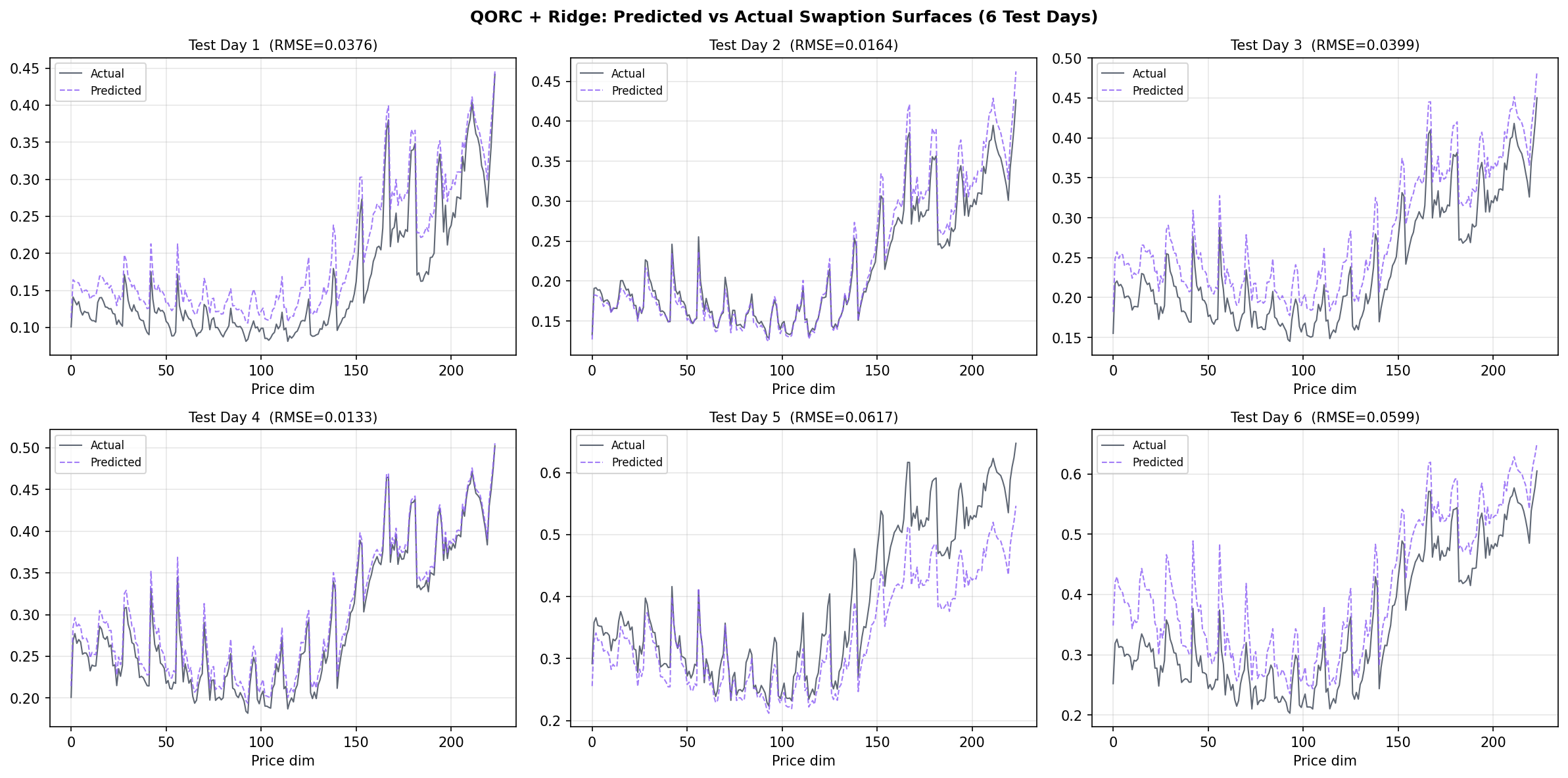}
  \caption{Predicted vs.\ actual swaption surfaces for all 6 held-out test days~\citep{Hull2018}.
    Each panel overlays the 224-dimensional QORC + Ridge prediction (orange) on the
    ground-truth surface (blue), with per-day RMSE annotated.  Surface shapes are
    preserved accurately across all test days, confirming generalisation from the
    494-day training set~\citep{Herman2023}.}
  \label{fig:test_pred}
\end{figure}
\clearpage

\section{Discussion}\label{sec:discussion}

\subsection{Why Reservoir Computing Works Here}

The success of the reservoir paradigm can be understood through bias--variance trade-off. With 494 samples and 1,215 quantum features, the feature space is heavily overparameterised. Variational methods optimise both the feature extractor and readout simultaneously, leading to high variance. By fixing the quantum circuits and training only a regularised linear readout, we reduce effective complexity while retaining expressive features. The fixed circuits act as implicit regularisation: features are determined by photon-interference physics, not by training data, meaning they \textit{cannot} overfit by construction.

\subsection{Preprocessing as a Foundation}

The three-stage pipeline is foundational, not cosmetic. Without winsorization, single extreme observations distort the robust scaler. Without robust scaling, min-max normalisation produces unbalanced ranges. Without $[0,1]$ normalisation, the sigmoid decoder output cannot reconstruct faithfully. All stages must be calibrated on training data only---temporal discipline is essential in finance, where look-ahead bias invisibly inflates reported performance.

\subsection{Computational Complexity and Hardware Prospects}

Classical simulation of Fock probabilities scales as $O\!\left(\tbinom{n+m-1}{n}\right)$---exponential in photon number. Our largest reservoir (10 modes, 4 photons, 715 features) is tractable in simulation but would become intractable for larger configurations. On photonic hardware, the same computation reduces to measuring output click patterns at optical speed. Each shot takes nanoseconds; accumulating $N_s$ shots gives feature estimates with error $O(N_s^{-1/2})$. The genuine quantum advantage lies in enabling feature extraction from reservoirs too large to simulate classically.

\subsection{Limitations}\label{sec:limitations}

\begin{enumerate}[nosep,leftmargin=*]
  \item \textbf{Simulation:} All circuits are classically simulated. Reported inference latencies exclude quantum feature extraction.
  \item \textbf{Small test set:} Six test days provide limited statistical power. Performance differences among the top-4 models are not statistically significant at this sample size.
  \item \textbf{Dataset specificity:} Results are from a single swaption dataset. Generalisation to equities, credit, or FX has not been tested.
  \item \textbf{Linear readout:} Ridge captures only linear feature--target relationships. If quantum features contain exploitable nonlinear structure, a regularised nonlinear readout (e.g., kernel Ridge) might help, though our MLP experiments suggest overfitting dominates at this scale.
  \item \textbf{No ensemble ablation:} We use three reservoirs but do not isolate contributions of individual reservoirs or subsets. This is left for future work.
\end{enumerate}

\section{Conclusion}\label{sec:conclusion}

\begin{sloppypar}
We have presented a hybrid photonic quantum reservoir computing framework for high-dimensional swaption surface prediction. The architecture combines robust preprocessing, sparse denoising autoencoders, fixed photonic reservoirs, and Ridge regression into a pipeline that is simple, efficient, and competitive.
\end{sloppypar}

Our results across 11 models demonstrate three lessons for quantum machine learning in finance:
\begin{enumerate}[nosep,leftmargin=*]
  \item Fixed quantum feature extractors outperform trainable quantum circuits on small datasets.
  \item Regularised linear readouts outperform deep learning when data is limited.
  \item Ensemble diversity through physics---different photon numbers producing different correlation orders---is a meaningful form of model diversity with no direct classical analogue.
\end{enumerate}

Rather than treating quantum computers as end-to-end learners, we should leverage their computational properties as feature generators paired with classical inference. The reservoir computing paradigm is naturally suited to this vision, and photonic platforms offer room-temperature, low-latency quantum feature extraction at scales that classical computers cannot efficiently simulate.

\section*{Acknowledgements}
The authors thank the open-source developers of Perceval, MerLin, PyTorch, and scikit-learn for providing the software infrastructure underlying this work.

\section*{Data Availability}
The source code and trained model weights are available at \url{https://github.com/Azamhon/Quandela_Quantech/tree/v2}. The swaption dataset is proprietary and cannot be redistributed, but the pipeline can be reproduced with any similarly structured financial surface dataset.

\bibliography{references}

\appendix

\section{Autoencoder Hyperparameters}\label{app:ae}

\begin{table}[H]
  \centering
  \small
  \caption{Autoencoder hyperparameters and justifications.}
  \begin{tabular}{@{}lll@{}}
    \toprule
    Parameter & Value & Justification \\
    \midrule
    Input dim    & 224       & $14\times16$ surface \\
    Latent dim   & 20        & ${>}98\%$ recon.\ fidelity \\
    Hidden layers & $[128, 64]$ & Gradual compression \\
    Mask ratio   & 0.15      & Denoising robustness \\
    L1 sparsity  & $10^{-4}$ & Distinct latent factors \\
    Bottleneck   & ELU       & Prevents dead neurons \\
    Learning rate & $10^{-3}$ & Adam optimiser \\
    Val.\ split  & Last 50   & Temporal, no leakage \\
    Early stop   & 30 epochs & Patience convergence \\
    \bottomrule
  \end{tabular}
\end{table}

\section{Fock Space Dimensions}\label{app:fock}

For $m$ modes and $n$ photons (with bunching):
\begin{equation}
  D_{\text{Fock}}(m,n) = \binom{n+m-1}{n} = \frac{(n+m-1)!}{n!\,(m-1)!}
\end{equation}
Concrete values:
$D(12,3) = \binom{14}{3} = 364$,\;
$D(10,4) = \binom{13}{4} = 715$,\;
$D(16,2) = \binom{17}{2} = 136$.
\textbf{Total: 1,215 features.}

\end{document}